\documentclass[preprint]{ptephy_v1}
\preprintnumber{KEK-CP-346, OU-HET-898}

\begin{document}

\title{Stochastic calculation of the Dirac spectrum on the lattice 
  and a determination of chiral condensate in 2+1-flavor QCD
\thanks{This paper is dedicated to the memory of Dr. Keisuke Jimmy
  Juge (1971-2016).}}

\author[1]{Guido Cossu}
\affil[1]{School of Physics and Astronomy, University of Edinburgh,
  EH9 3JZ, Edinburgh, United Kingdom}

\author[2]{Hidenori Fukaya}
\affil[2]{Department of Physics, Osaka University, Toyonaka 560-0043, Japan}

\author[3,4]{Shoji Hashimoto}
\affil{High Energy Accelerator Research Organization (KEK), 
  Tsukuba 305-0801, Japan}

\affil[4]{School of High Energy Accelerator Science,
  SOKENDAI (The Graduate University for Advanced Studies),
  Tsukuba 305-0801, Japan}

\author[3,4]{Takashi Kaneko}

\author[3]{Jun-Ichi Noaki}


\begin{abstract}
  We compute the chiral condensate in 2+1-flavor QCD through the
  spectrum of low-lying eigenmodes of Dirac operator.
  The number of eigenvalues of the Dirac operator is evaluated using
  a stochastic method with an eigenvalue filtering technique
  on the background gauge configurations generated by lattice QCD
  simulations including the effects of dynamical up, down and strange
  quarks described by the M\"obius domain-wall fermion formulation.
  The low-lying spectrum is related to the chiral condensate,
  which is one of the leading order low-energy constants in chiral
  effective theory, as dictated by the Banks-Casher relation.
  The spectrum shape and its dependence on the sea quark masses 
  calculated in numerical simulations 
  are consistent with the expectation from one-loop chiral
  perturbation theory.
  After taking the chiral limit as well as the continuum limit using
  the data at three lattice spacings ranging 0.080-0.045~fm, we obtain
  $\Sigma^{1/3}(\mathrm{2~GeV})$ = 270.0(4.9)~MeV, 
  with the error combining those
  from statistical and from various sources of systematic errors.
  Finite volume effect is confirmed to be under control by a direct
  comparison of the results from two different volumes at the lightest
  available sea quarks corresponding to 230~MeV pions.
\end{abstract}

\date{\today}

\maketitle

\section{Introduction}
Spectrum $\rho(\lambda)$ of the eigenvalues $\lambda$ of the Dirac 
operator $D$ in Quantum Chromodynamics (QCD) reflects the properties
of background gauge field.
At zero temperature, pairs of quark and antiquark condense in the
vacuum as represented by the Banks-Casher relation 
$\rho(0)=\Sigma/\pi$ \cite{Banks:1979yr}, which is valid in the
thermodynamical limit, {\it i.e.} massless quark limit after taking an
infinite volume limit.
In other words, the density of near-zero eigenvalues of the Dirac
operator is related to the chiral condensate $\Sigma$, which is an
order parameter of spontaneous chiral symmetry breaking in QCD.
In the chiral effective theory, for which pions play the role of
effective degrees of freedom of QCD at low energy, 
the chiral condensate $\Sigma$ and pion decay constant $F_\pi$ are the
most fundamental parameters appearing at the lowest order in an
expansion in terms of pion mass and momenta.
The QCD Dirac spectrum can thus be related to physical observables
involving pions at low energy.

The chiral effective theory predicts the functional form of
$\rho(\lambda)$ in the low energy regime.
In the limit of infinite volume, the slope of $\rho(\lambda)$ at
$\lambda=0$ was calculated including the loop effect of pions
\cite{Smilga:1993in}, and the dependence on the number of dynamical
quark flavors was predicted.
In a finite volume, the lowest end of the spectrum is largely affected
and exact zero-modes play a special role.
Such system is related to the chiral Random Matrix Theory (RMT),
with which the distribution of individual eigenvalue can be calculated
\cite{Osborn:1998qb}. 
(For more results, see a recent review article
\cite{Akemann:2016keq}.)
The most elaborate calculation to date includes finite volume and
finite quark mass corrections in a systematic expansion
\cite{Damgaard:2008zs}.

In lattice gauge theory calculations, the spectral density has so far
been calculated by direct computation of the low-lying eigenvalues or
by stochastic estimates of the mode number below some value
\cite{Giusti:2008vb}.
The direct computation of individual eigenvalues has an advantage of
allowing a comparison of the microscopic distribution with that
predicted by chiral RMT.
Even with a few lowest eigenvalues, one can then extract $\Sigma$
assuming the correspondence between the chiral effective theory and
the random matrix theory.
In our previous works using the overlap fermion formulation, 
we studied the quark mass and volume dependence of the eigenvalue
distribution and extracted the value of $\Sigma$ in 2-flavor
\cite{Fukaya:2007fb,Fukaya:2007yv} 
and 2+1-flavor QCD
\cite{Fukaya:2009fh,Fukaya:2010na}.
Since the overlap fermion preserves exact chiral symmetry, the
smallest eigenvalues satisfy the relations derived from chiral
symmetry, and
the correspondence between the non-perturbative lattice calculation
and the analytic prediction of the effective theory and chiral RMT
\cite{Damgaard:1997ye,Damgaard:2000ah} has been precisely
established. 

In order to achieve precise calculation of the physical value of
$\Sigma$, on the other hand, the direct eigenvalue calculation with
the exactly chiral fermion formulation is computationally too
expensive. 
Finite volume effect and discretization effect are best controlled by
calculating on sufficiently large and fine lattices.
The number of relevant low-lying eigenvalues to be calculated grows as
the (four-dimensional) volume $V$, and the computation of individual
eigenvalues rapidly becomes impractical.
The stochastic estimate introduced in \cite{Giusti:2008vb} offers an
alternative method in such situations.
The method has been successfully applied to extract $\Sigma$ in 2-
and/or 2+1-flavor QCD with Wilson \cite{Engel:2014cka,Engel:2014eea}  
and twisted-mass \cite{Cichy:2013gja} fermion formulations.

In this work we use a slightly different implementation of the
stochastic estimate.
It is based on a filtering of eigenvalues in a given interval
\cite{DiNapole:2013}.
The method allows us to estimate the number of eigenvalues in any
interval once the necessary coefficients have been calculated.
We use the domain-wall fermion formulation, with which chiral symmetry
can be maintained at the level that the effective residual quark mass
is of order of 1~MeV.
We design the eigenvalue filtering such that the number of
eigenvalues in a bin of 5~MeV or larger is counted 
and the possible effect of the residual chiral symmetry violation is
harmless.

We calculate the eigenvalue spectrum on the lattices generated with
2+1 flavors of light sea quarks described by the M\"obius domain-wall
fermion.
Sea quark masses in the simulations correspond to the pion mass in the
range of 230--500~MeV.
Physical volume is sufficiently large, $L\sim$ 2.6~fm or larger, in
order to safely neglect the effect of finite volume which affects the
lowest eigenvalues of order $\lambda\sim 1/(\Sigma V)$
($\sim$ 1--2~MeV) most strongly 
while the number of eigenvalues below 10--20~MeV is little affected.
The finite volume effect due to the loop effects of light pions 
is suppressed as $\exp(-M_\pi L)$, and is sufficiently small on our
lattices satisfying $M_\pi L> 4$.

Our lattice ensembles are in a range of lattice spacing $a$ 
between 0.080--0.044~fm.
The corresponding lattice cutoff $a^{-1}$ ranges between 2.45~GeV and
4.50~GeV.  
On these fine lattices, the discretization effects for the near-zero 
eigenvalues of order 10~MeV should be negligible.
Indeed, we found that the scaling violation is consistent with zero
for the spectral function.

The M\"obius domain-wall fermion is an (approximate) implementation of
the Ginsparg-Wilson relation \cite{Ginsparg:1981bj}.
The residual mass with our parameter choices is $O(\mbox{1~MeV})$ or
less strongly depending on the lattice spacing, and its effect on the
calculation of the eigenvalue spectral density is minor.

Using these data sets we obtain the spectral density, which we then
fit with the formula predicted by the chiral effective theory to
obtain the value of chiral condensate $\Sigma$ in the chiral limit of
up and down quarks.

The rest of the paper is organized as follows.
In Section~\ref{sec:stoch-estim-eigenv} we review the method of
the eigenvalue filtering and the stochastic eigenvalue counting.
Section~\ref{sec:domain-wall-fermion} summarizes the lattice
fermion formulation, which is followed by the details of our data sets
in Section~\ref{sec:lattice_ensembles}.
The spectral function in the entire range of eigenvalues is shown in
the plots given in Section~\ref{sec:spectral_function}.
We then focus on the low-lying eigenvalue spectrum to extract the
low-energy constants including the chiral condensate using the chiral
perturbation theory, as described in Section~\ref{sec:chPT_analysis}.
Our conclusion is in Section~\ref{sec:conclusion}.
A preliminary report of this work is found in \cite{Cossu:2016yzp}.

\section{Stochastic estimate of eigenvalue count}
\label{sec:stoch-estim-eigenv}
We review the method to evaluate the eigenvalue count of a hermitian
matrix in a given interval.
More details are described in \cite{DiNapole:2013}.
In the lattice gauge theory calculations, the method is 
introduced recently in \cite{Fodor:2016hke}.

Let $A$ be a hermitian matrix and assume that its eigenvalues are
distributed in the range $[-1,1]$.
If not, we can easily rescale the matrix by a linear transformation.
We aim at calculating the number of eigenvalues of this matrix in a
given interval $[s,t]$.
By introducing a step function $h(A)$ that has a value 1 only
in the interval $[s,t]$ and zero elsewhere,
the number of eigenvalues is written as 
$n[s,t]=\mathrm{Tr}\,h(A)$.
Then, introducing $N_v$ Gaussian random noise vectors $\xi_k$ with a
normalization 
$(1/N_v)\sum_{k=1}^{N_v}\xi_k^\dagger\xi_k=12V$
in the limit of large $N_v$,
one may evaluate $n[s,t]$ as 
\begin{equation}
  \label{eq:n}
  n[s,t] = \frac{1}{N_v}\sum_{k=1}^{N_v}
  \xi_k^\dagger h(A) \xi_k
\end{equation}
in the limit of large $N_v$.
This evaluation can be promoted to the ensemble average as
\begin{equation}
  \label{eq:n_ave}
  \bar{n}[s,t] = \frac{1}{N_v}\sum_{k=1}^{N_v}
  \langle\xi_k^\dagger h(A) \xi_k\rangle,
\end{equation}
where $\langle\cdots\rangle$ represents an average over Monte Carlo
samples, or the gauge configurations.
With sufficiently large number of gauge configurations, we may even
take $N_v=1$ to obtain a statistically significant signal.

The discrete function $h(A)$ may be constructed approximately using a
polynomial function even when the matrix $A$ is large.
The best approximation of $h(x)$ in the sense of min-max 
(smallest maximum deviation) in the interval $x\in[-1,1]$
achieved within a given computation cost is the Chebyshev
approximation using the Chebyshev polynomial $T_j(x)$.
Explicitly, we may write
\begin{equation}
  \label{eq:Chebyshev expansion}
  h(x) \simeq \sum_{j=0}^p \gamma_j T_j(x),
\end{equation}
with coefficients $\gamma_j$, which can be calculated as a function of
$s$ and $t$.
See equation (7) of \cite{DiNapole:2013}, which is reproduced below
for convenience:
\begin{equation}
  \gamma_j = \left\{
    \begin{array}{ll}
      \displaystyle
      \frac{1}{\pi}\left(\arccos(s)-\arccos(t)\right) & \;\; j=0,\\
      \displaystyle
      \frac{2}{\pi}\left(
      \frac{\sin(j\arccos(s))-\sin(j\arccos(t))}{j}
      \right) & \;\; j>0.
    \end{array}
  \right.
\end{equation}
In order to suppress a strong oscillation emerging with this
approximation, 
the so-called Jackson damping factor $g_j^p$ is introduced,
sacrificing the ``best'' approximation. 
An explicit formula, equation (10) of \cite{DiNapole:2013}, is 
\begin{equation}
  g_j^p = \frac{
    \left(1-\frac{j}{p+2}\right)
    \sin\alpha_p \cos(j\alpha_p) +
    \frac{1}{p+2}
    \cos\alpha_p \sin(j\alpha_p)}{\sin\alpha_p},
\end{equation}
where $\alpha_p=\pi/(p+2)$.

The formula (\ref{eq:Chebyshev expansion}) can then be modified as
\begin{equation}
  \label{eq:Chebyshev expansion Jackson}
  h(x) \simeq \sum_{j=0}^p g_j^p \gamma_j T_j(x).
\end{equation}
Using this form, the stochastic estimate of (\ref{eq:n_ave}) can be
approximated as
\begin{equation}
  \label{eq:n_ave_Cheby}
  \bar{n}[s,t] \simeq \frac{1}{N_v}\sum_{k=1}^{N_v}
  \left[\sum_{j=0}^p g_j^p\gamma_j
    \langle\xi_k^\dagger T_j(A) \xi_k\rangle
  \right].
\end{equation}
This approximation is convenient, because one can obtain the
eigenvalue count in any range $[s,t]$ once we have the set of
measurements for $\langle\xi_k^\dagger T_j(A)\xi_k\rangle$.

The Chebyshev polynomial is constructed using the recursion relation:
$T_0(x)=1$, $T_1(x)=x$ and
\begin{equation}
  \label{eq:recursion}
  T_j(x) = 2x T_{j-1}(x) - T_{j-2}.
\end{equation}
There is also an useful formula,
$2T_m(x)T_n(x)=T_{m+n}(x)+T_{|m-n|}(x)$, which in particular reads
\begin{equation}
  \left\{
    \begin{array}{ccc}
      T_{2n-1}(x) & = & 2 T_{n-1}(x)T_n(x) - T_1(x)\\
      T_{2n}(x) & = & 2 T_n^2(x) - T_0(x)
    \end{array}
  \right..
\end{equation}
One can then apply $A$ on $\xi_k$ repeatedly to obtain 
$\xi_k^\dagger T_j(A)\xi_k$.
Note that the $2n$-th order is obtained from 
$(T_n(A)\xi)^\dagger (T_n(A)\xi)$
using the formula above.
One therefore needs $n$ multiplication of $A$ to obtain the order of
polynomial $p=2n$.

The accuracy of the approximation depends on the order of the
polynomial $p$. 
The size of error is discussed in the next section for the 
application to the spectral function of the domain-wall fermion Dirac
operator.

\section{Domain-wall Dirac operator}
\label{sec:domain-wall-fermion}
In this work we utilize the M\"obius domain-wall fermion formulation
\cite{Brower:2012vk} to define the Dirac operator on the lattice.
It is a generalization of the domain-wall fermion
\cite{Kaplan:1992bt,Shamir:1993zy} 
introduced to achieve better chiral symmetry within a given
computational cost.
In this fermion formulation, the fermion field is defined on a
five-dimensional (5D) lattice, and a four-dimensional (4D) fermion
emerges on the 4D surfaces of the 5D space.
The fermion modes of right-handed and left-handed chiralities localize
on the opposite 4D surfaces, and thus chiral fermion is realized with
exponentially suppressed violation as a function of the extent in the
fifth direction $L_s$.

The effective 4D Dirac operator $D^{(4)}$ is constructed combining the
5D Dirac operator $D_{DW}^{(5)}(m)$ with a fermion mass $m$ as
\cite{Brower:2012vk}
\begin{equation}
  \label{eq:D4}
  D^{(4)} = \left[
    {\cal P}^{-1} (D_{GDW}^{(5)}(1))^{-1} 
    D_{GDW}^{(5)}(0) {\cal P}
    \right]_{11}.
\end{equation}
Here ${\cal P}$ is a certain permutation operator acting on the fifth 
coordinate $s$ designed to move the physical surface modes (both 
left-handed and right-handed) to the slice of $s=1$.
The suffix ``11'' then means to extract that 4D slice.
The term $(D_{GDW}^{(5)}(1))^{-1}$ implies an introduction of a
Pauli-Villars field, which cancels unnecessary 5D modes in the
ultraviolet limit.

The 4D operator $D^{(4)}$ approximately satisfies the 
Ginsparg-Wilson relation \cite{Ginsparg:1981bj} 
\begin{equation}
  \label{eq:GW}
  D^{(4)}\gamma_5+\gamma_5 D^{(4)} = 2 D^{(4)}\gamma_5 D^{(4)},
\end{equation}
and the eigenvalues of the hermitian operator 
$D^{(4)\dagger} D^{(4)}$ are constrained in the
range $[0,1]$.
In order to apply the eigenvalue filtering method described in the
previous section, we therefore define
\begin{equation}
  \label{eq:DdagD}
  A = 2 D^{(4)\dagger} D^{(4)} - 1,
\end{equation}
such that $A$ has eigenvalues between $-1$ and $1$.

The low upper limit (=1) of the eigenvalue of $D^{(4)\dagger} D^{(4)}$
is one of the advantages of using the domain-wall fermion.
With the Wilson fermion formulation, for instance, the highest
eigenvalue is $8^2=64$ (or slightly less for interacting cases) and
one has to shrink the whole eigenvalue range by multiplying a factor
$\sim$~30 to fit in $[-1,1]$ when we map the Wilson operator on $A$ as
in (\ref{eq:DdagD}).
The target eigenvalue interval is then much narrower for $A$, and one
needs larger polynomial order $p$ to obtain the same level of accuracy.
Although the numerical cost is higher for the domain-wall fermion due
to the inversion of the Pauli-Villars operator for each application of
$D^{(4)}$, the difference of the entire eigenvalue range nearly
compensates the cost compared to the Wilson fermion.

An eigenvalue $a^2\lambda_{D^\dagger D}$ of $D^{(4)\dagger} D^{(4)}$ can
be related to that of $D^{(4)}$ assuming the Ginsparg-Wilson relation
(\ref{eq:GW}) as well as the $\gamma_5$-hermiticity property 
$D^{(4)\dagger}=\gamma_5D^{(4)}\gamma_5$.
The relation (\ref{eq:GW}) is slightly violated in the actual
implementation, and the associated error is discussed later.
The eigenvalues $\lambda_D$ of $D^{(4)}$ lie on a circle on the
complex plane to satisfy $|a\lambda_D-1/2|=1/2$.
We project them to the imaginary axis to obtain the 
{\it continuum-like} eigenvalue $\lambda$ as
\begin{equation}
  \label{eq:proj}
  a\lambda \equiv 
  \sqrt{\frac{a^2\lambda_{D^\dagger D}}{1-a^2\lambda_{D^\dagger D}}}.
\end{equation}
This is a convention, and other definitions such as
$a\lambda=|a\lambda_D|$ are equally valid up to the discretization
effect of $O(a^2)$.
For the low-lying modes below 20~MeV, which are the eigenmodes we use
to extract the chiral condensate, 
the discretization error of
$O(a^2)$ is expected to be very small.

\begin{figure}[tbp]
  \centering
  \includegraphics[width=10cm,clip]{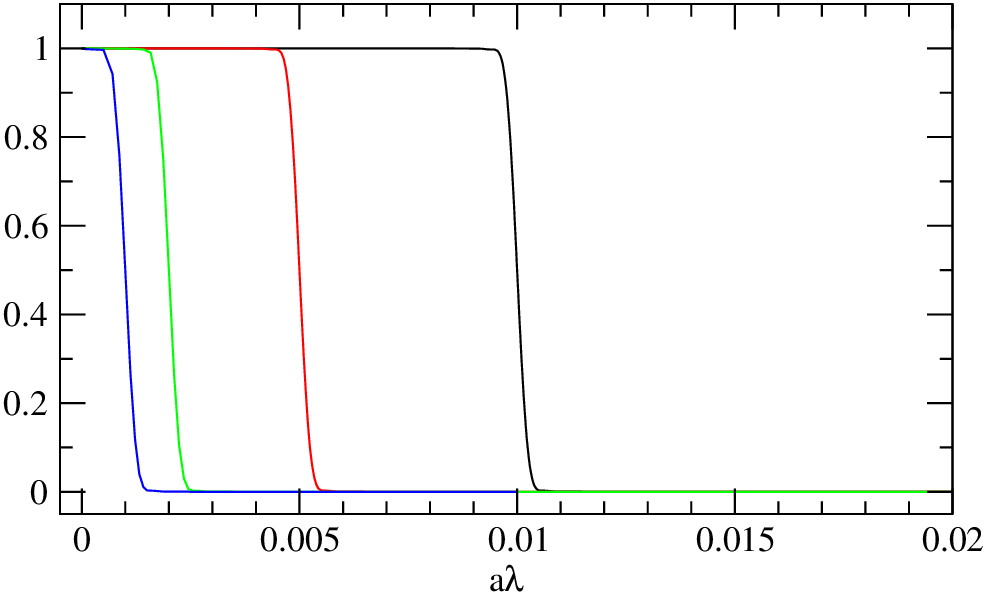}
  \caption{
    Approximate step function for the lowest bin 
    $a\lambda=[0,a\delta]$.
    The cases of $a\delta$ = 0.01, 0.005, 0.002 and 0.001 are
    plotted.
    The degree of polynomial is $p$ = 8000.
  }
  \label{fig:step}
\end{figure}

Examples of the filtering function are shown in
Figure~\ref{fig:step}
for the order of polynomial $p$ = 8000.
Here, the Dirac eigenvalue $a\lambda$ as defined in (\ref{eq:proj}) is
taken on the horizontal axis.
The plot shows the function to extract the count in the lowest bin
$[0,a\delta]$ of bin size $a\delta$ = 0.01, 0.005, 0.002 and 0.001.
The approximation of the step function is very precise except for the
region close to the threshold $a\lambda=a\delta$.
The width where the function varies is nearly independent of
$a\delta$, and as a result, the relative error of the approximation is
smaller for larger bin sizes.
Note that the lowest eigenvalue is the worst case, because it is
mapped onto a narrow bin of size $2(a\delta)^2$ of $A$.

\begin{figure}[tbp]
  \centering
  \includegraphics[width=10cm,clip]{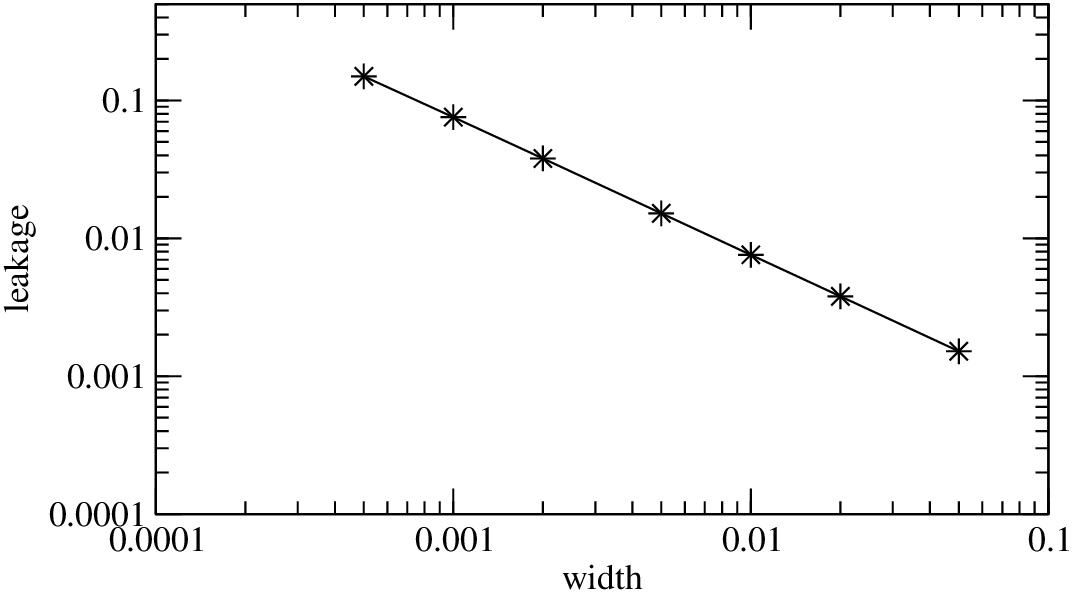}
  \caption{
    Fractional leakage from the lowest bin $\lambda=[0,\delta]$ to the 
    neighboring bin.
    The fraction of the leakage is plotted as a function of the bin
    size $a\delta$.
    The case of $p$ = 8000 is plotted.
  }
  \label{fig:leak}
\end{figure}

In order to quantify the size of the error in filtering, we calculate
a fraction of leakage from the lowest bin $[0,a\delta]$ to the
neighboring bin.
It is defined as an integral of the filtering function from $a\delta$
to infinity, which should vanish for the exact step function.
The leakage is equal to the deficit in the bin of interest 
$[0,a\delta]$.
Figure~\ref{fig:leak} shows the leakage for various widths $a\delta$.
The relative error increases for smaller $a\delta$ as an inverse power
$1/a\delta$. 
If we allow an 1\% error for the calculation of the spectral
function, we may take $a\delta$ to be 0.005 when $p$ = 8000.
This bin size corresponds to 12~MeV on our coarsest lattice.
On finer lattices we take larger values of $p$ so that the error 
with a fixed $\delta$, which implies a smaller $a\delta$ on a finer
lattice, is not larger than 0.005.
Since the deficit is largely compensated by the leakage from the
neighboring bin when the spectral function is nearly constant as it is
the case for zero temperature QCD, the actual error would be much
smaller than this naive estimate.

According to the general theory of the Chebyshev approximation, the
error as measured by the $L_2$ norm scales as $1/\sqrt{p}$
\cite{DiNapole:2013}.
With the Jackson damping factor implemented in this work, this bound
does not apply, but an actual calculation as outlined above indicates 
that the leakage decreases as $1/p$.
This determines the computational cost when one wants to improve the
precision using this method.

\section{Lattice ensembles}
\label{sec:lattice_ensembles}

\begin{table}[tbp]
  \centering
  \begin{tabular}{ccccccccc}
    \hline
    $\beta$ & $a^-1$ & $L/a$ & $am_{ud}$ & $am_s$ & $m_\pi$ & $p$ & $N_{meas}$\\
    & [GeV] & & & & [MeV] & & \\
    \hline
    4.17 & 2.453(4) & 32 & 0.019 & 0.030 & 498.0(0.7) & 8,000 & 100\\
         &          &    & 0.012 &       & 396.8(0.7) &       & 100\\
         &          &    & 0.007 &       & 309.8(1.0) &       & 100\\
         &          &    & 0.019 & 0.040 & 498.7(0.7) & 8,000 & 100\\
         &          &    & 0.012 &       & 399.0(0.8) &       & 100\\
         &          &    & 0.007 &       & 309.2(1.0) &       & 100\\
         &          &    & 0.0035 &      & 229.8(1.1) &       & 100\\
         &          & 48 & $\uparrow$ &  & 225.8(0.3) & 9,000 & 100\\
    \hline
    4.35 & 3.610(9) & 48 & 0.0120 & 0.0180 & 498.5(0.9) & 16,000 & 50\\
         &          &    & 0.0080 &        & 407.0(1.2) &        & 50\\
         &          &    & 0.0042 &        & 295.9(1.2) &        & 50\\
         &          &    & 0.0120 & 0.0250 & 500.7(1.0) & 16,000 & 50\\
         &          &    & 0.0080 &        & 407.8(1.0) &        & 50\\
         &          &    & 0.0042 &        & 299.9(1.2) &        & 50\\
    \hline
    4.47 & 4.496(9) & 64 & 0.0030 & 0.0150 & 284.2(0.7) & 15,000 & 40\\
    \hline
  \end{tabular}
  \caption{
    Lattice ensembles used in the eigenvalue spectrum calculation.
    Spatial lattice size $L/a$ and sea quark masses $am_{ud}$, $am_s$
    are listed in the lattice unit.
    $p$ is the order of the Chebyshev polynomial, and $N_{meas}$ is
    the number of measurements.
    Empty entries are the same as the ones in the previous line.
  }
  \label{tab:ensemble}
\end{table}

We calculate the spectral function at three $\beta$ values on 15 gauge
ensembles in total, generated with 2+1 flavors of sea quarks
\cite{Noaki:2014sda}, as listed in Table~\ref{tab:ensemble}.
The formulation for the sea quarks is the M\"obius domain-wall
fermion, which is the same for the lattice Dirac operator used in the
eigenvalue counting.
The gauge action is tree-level Symanzik improved, and we apply the
stout link smearing \cite{Morningstar:2003gk} three times
for the link variables entering the definition
of the fermionic operators.

The lattice spacings determined through the Wilson flow scale 
$t_0$ are 0.0803(1), 0.0546(1) and 0.0438(1)~fm at 
$\beta$ = 4.17, 4.35, 4.47, respectively, 
where we report only the statistical error.
We chose the input $t_0^{1/2}$ = 0.1465(21)(13)~fm from 
\cite{Borsanyi:2012zs}.
The error on this input value is taken into account as one of the
sources of systematic error.

Except for the finest lattice ($\beta$ = 4.47), we generated lattices
at several values of $(am_{ud},am_s)$, combinations of the up/down
and strange quark masses.
Corresponding pion mass $m_\pi$ covers the range 
between 230 and 500~MeV.
Two strange quark masses sandwich its physical value.
The finest lattice at $\beta$ = 4.47 is available only at one
combination of sea quark masses.
The corresponding pion mass is about 280~MeV.

The spatial extent of the lattice $L/a$ is chosen such that the
physical size $L$ is kept constant around 2.6--2.8~fm.
The measure of the finite volume effect $m_\pi L$ is larger than 3.9
for all ensembles except for the one of the lightest sea quark mass
($am_{ud}$ = 0.0035) on the $L/a=32$ lattice.
For this parameter we prepare a lattice ensemble of larger volume,
$L/a=48$, in order to examine the finite volume effect.
On this larger lattice, $m_\pi L$ = 4.4.
The results from the $L/a=32$ lattice at this parameter are used only
to investigate the finite volume effect and not included in the final
analysis of the chiral condensate.
The temporal size $T$ is always twice as large as the spatial size
$L$. 

For each ensemble we run a Hybrid Monte Carlo simulation for 10,000
molecular-dynamics trajectories, out of which we chose (equally
separated) $N_{meas}$ = 40--100 gauge configurations for the
calculation of the spectral function.

The M\"obius domain-wall fermion is defined on a 5D lattice.
The extent in the fifth dimension $L_s$ is chosen such that the
violation of the Ginsparg-Wilson relation is sufficiently small.
By taking $L_s$ = 12 on the coarsest lattice at $\beta$ = 4.17
we confirm that the residual mass is roughly 1~MeV
\cite{Hashimoto:2014gta}.
On the finer lattices at $\beta$ = 4.35 and 4.47, we take $L_s$ = 8
and the residual mass is much smaller: 0.2~MeV at $\beta$ = 4.35 and 
$<0.1$~MeV at $\beta$ = 4.47.
These small but non-zero residual mass may distort the low-lying Dirac 
spectrum. 
With the bin size we chose to count the eigenvalues, such effect
would be minor; we eventually eliminate the associated error by taking
the continuum limit using the three lattice spacings we prepared.

The same set of ensembles is used for a wide variety of applications
including 
a determination of non-perturbative renormalization constant
\cite{Tomii:2016xiv},
a determination of the charm quark mass from temporal moments of
charmonium correlator
\cite{Nakayama:2016atf},
a calculation of the $\eta'$ meson mass through a gluonic observable 
\cite{Fukaya:2015ara}, and
a calculation of $D_{(s)}$ meson decay constant
\cite{Fahy:2015xka}.
Numerical calculations of the projects are performed using the code
set IroIro++ \cite{Cossu:2013ola}.

\section{Spectral function: overview}
\label{sec:spectral_function}
First, we demonstrate how the eigenvalue filtering method works by
showing the results on our coarsest lattices, {\it i.e.}
$32^3\times 64$ lattices at $\beta=4.17$.
The sea quark masses are
 $(am_{ud},am_s)$ = (0.0035,0.040),
(0.007, 0.040), (0.012, 0.040), (0.019, 0.040).
The corresponding pion mass ranges between 230~MeV and 500~MeV.

Averaging over 50 gauge configurations each with only one noise per 
configuration, we calculated $\langle\xi_k^\dagger T_j(A)\xi_k\rangle$.
The mode number $\bar{n}[s,t]$ is then evaluated by summing over $j$
from 0 to $p$ as (\ref{eq:Chebyshev expansion Jackson}).
The spectral density is obtained with an appropriate normalization,
\begin{equation}
  \label{eq:spectral_density}
  a^3\rho(\lambda;\delta) 
  = \frac{1}{2V/a^4} \frac{\bar{n}[s,t]}{a\delta},
\end{equation}
where 
$a\lambda=s^{1/2}/(1-s)$ and 
$a(\lambda+\delta)=t^{1/2}/(1-t)$.
The factor 2 in the denominator of (\ref{eq:spectral_density})
reflects the pairing of the eigenvalues, {\it i.e.} $\pm i\lambda$.

\begin{figure}[tb]
  \centering
  \includegraphics[width=10cm,clip]{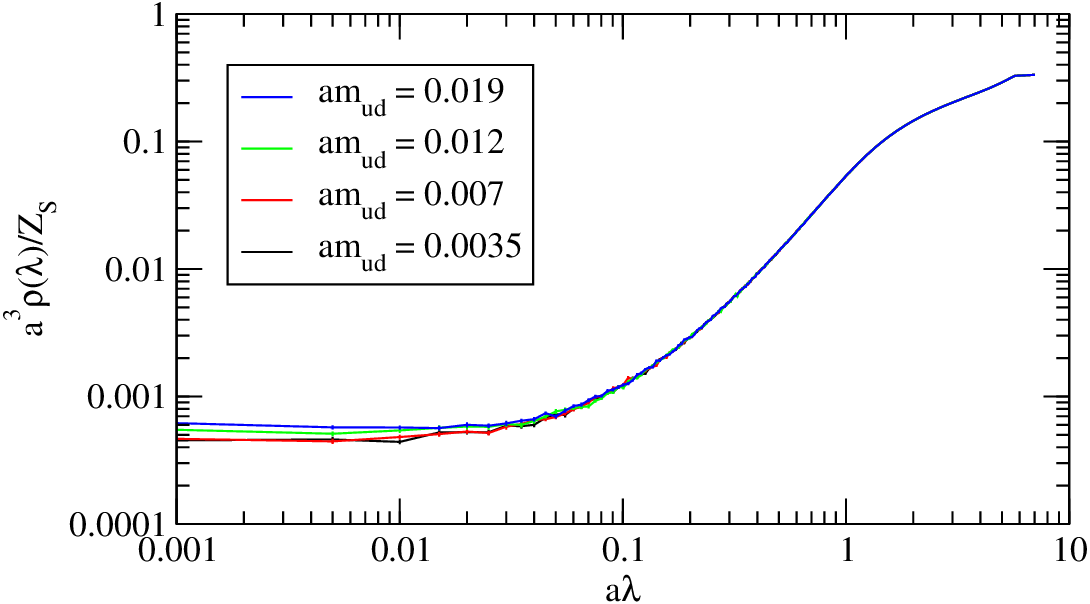}
  \caption{
    Spectral function in the entire range of $a\lambda$.
    Both the horizontal and vertical axes are logarithmic. 
    Data obtained at $\beta$ = 4.17 at four values of light quark mass
    $am_{ud}$.
  }
  \label{fig:spect_log}
\end{figure}

Figure~\ref{fig:spect_log} shows the spectral density 
$a^3\rho(\lambda)$ in the whole range of $a\lambda$.
The bin size is $a\delta$ = 0.005.
One can clearly observe that the spectrum starts from a tiny constant
at $\lambda\simeq 0$ and increases towards higher eigenvalues.
The near-zero modes show some dependence on the sea quark mass (see
below), but the high modes are nearly independent of the sea quark
mass. 

The increase towards the perturbative regime at high $a\lambda$ 
is qualitatively consistent with the free-theory scaling 
$\sim\lambda^3$, but saturates at around $a\lambda\sim 1$ due to the
discretization effect.

\begin{figure}[tbp]
  \centering
  \includegraphics[width=7.5cm,clip]{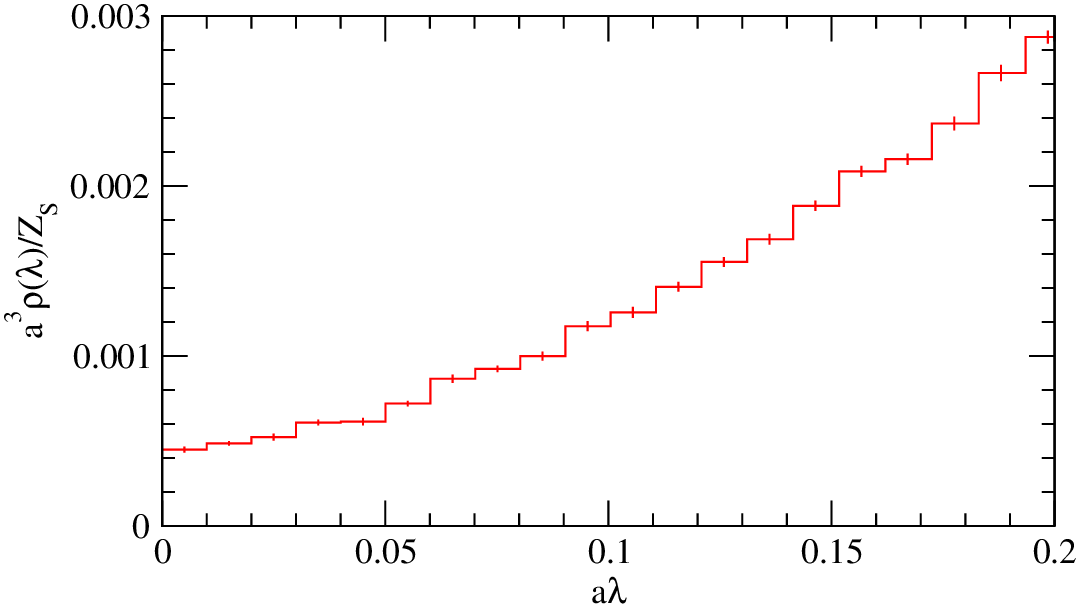}
  \includegraphics[width=7.5cm,clip]{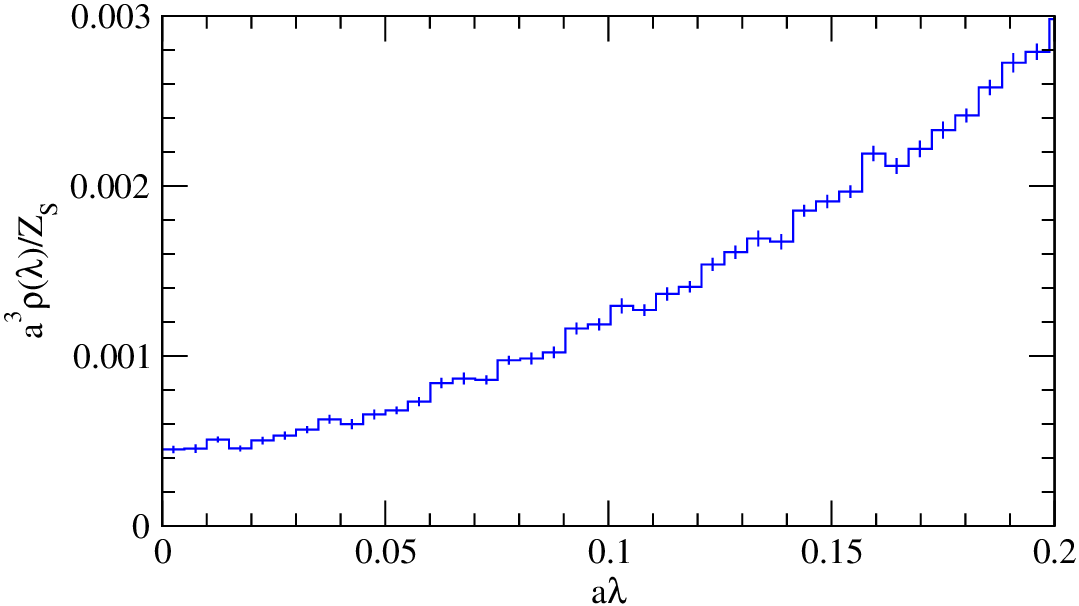}
  \includegraphics[width=7.5cm,clip]{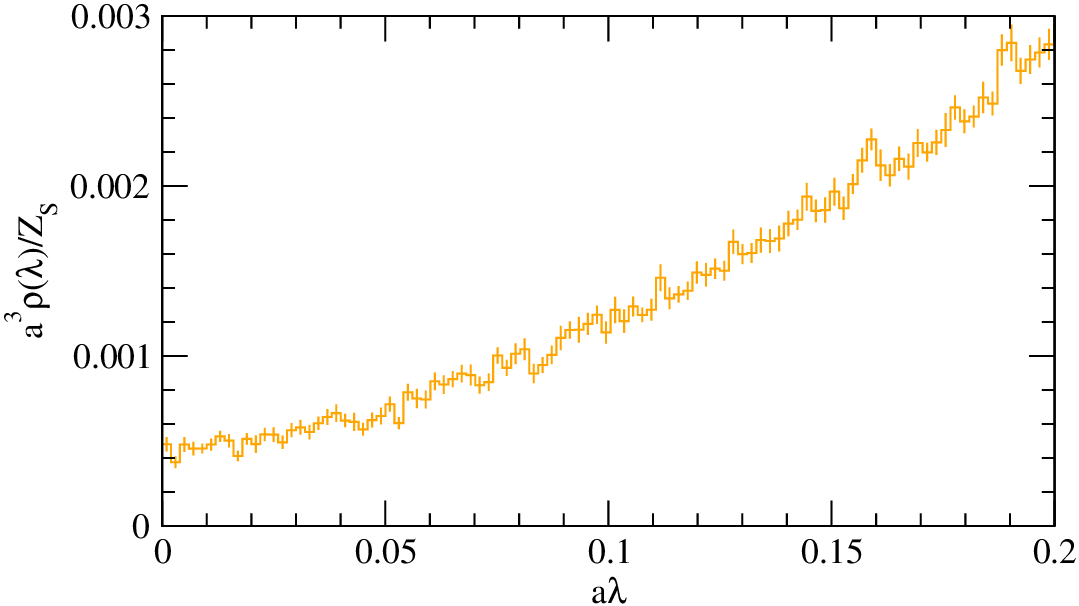}
  \includegraphics[width=7.5cm,clip]{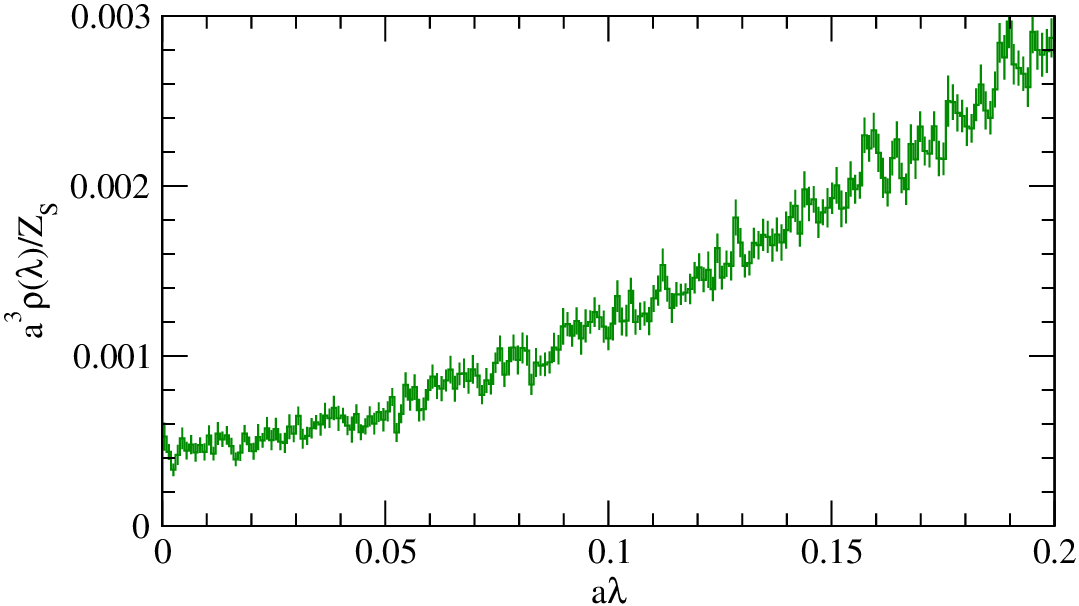}
  \caption{
    Spectral function in the low-lying region.
    The data at $\beta=4.17$ and $(am_{ud},am_s)$ = (0.007, 0.030).
    Results with different bin sizes are shown: 
    $a\delta$ = 0.01, 0.005, 0.002, and 0.001 
    from top left to bottom right.
  }
  \label{fig:spect_bins}
\end{figure}

Figure~\ref{fig:spect_bins} shows the spectral function in the
low-lying regime. 
The data at $(am_{ud},am_s)$ = (0.007, 0.030) are shown.
Results of different bin sizes ($a\delta$ = 0.01, 0.005, 0.002, 0.001)
are plotted.
We find that they are consistent within the statistical errors.
The statistical error is larger for smaller bins since the number of
eigenvalues in each bin is fewer.

\begin{figure}[tbp]
  \centering
  \includegraphics[width=10cm,clip]{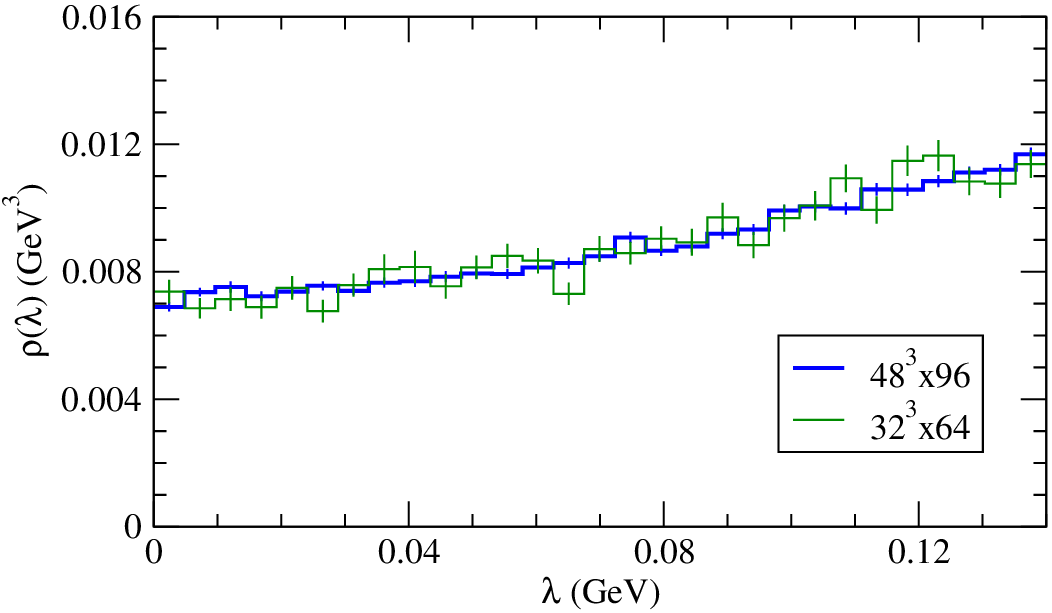}
  \caption{
    Volume scaling of the spectral function $\rho(\lambda)$.
    The data on the coarsest lattice ($\beta$ = 4.17) at two different
    volumes $48^3\times 96$ (thick line) and $32^3\times 64$ (thin) at
    $am_{ud}$ = 0.0035 and $am_s$ = 0.040.
  }
  \label{fig:spect_b417_vol}
\end{figure}

The volume scaling of the spectral function is demonstrated in Figure~\ref{fig:spect_b417_vol}.
For the lightest pion ($m_\pi\simeq$ 230~MeV), there are data on two
volumes $32^3\times 64$ and $48^3\times 96$ available.
We calculate the spectral density on both lattices with exactly the
same method.
The results are consistent with each other within the statistical
error, which is about 5\% on the $32^3\times 64$ lattice.
The statistical error is smaller, about 2\%, on the larger volume
since the number of eigenvalues in a given bin is proportional to the
physical volume $V$.

\section{Analysis with chiral perturbation theory}
\label{sec:chPT_analysis}
The Banks-Casher relation $\rho(0)=\Sigma/\pi$ is valid only in the
chiral limit after taking the infinite volume limit.
Therefore, the effects of finite sea quark masses and finite volume
need to be taken into account in the analysis.
We use the functional form predicted by the chiral effective theory to
analyze the quark mass dependence.
The finite volume effect is also estimated within the same framework,
but it turned out to be negligible in our setup as discussed below.

The analytic calculation is available at the one-loop order of chiral
perturbation theory ($\chi$PT), which is valid at the leading
non-trivial order of finite quark mass correction, {\it i.e.} of order
$m_\pi^2/(4\pi F_\pi)^2$.
The formula is concisely written in the form 
\cite{Damgaard:2008zs} (see also \cite{Fukaya:2010na})
\begin{equation}
  \label{eq:rho}
  \rho(\lambda) = \frac{\Sigma}{\pi} \left[
    1 - \frac{1}{F^2} \left(
      \sum_i\mathrm{Re}\Delta(0,M_{vi}^2) 
      - \mathrm{Re} G(0,M_{vv}^2,M_{vv}^2)
      - 16 L_6 \sum_i M_{ii}^2 \right)
  \right]_{m_v=i\lambda}, 
\end{equation}
where the chiral condensate $\Sigma$ and pion decay constant $F$ are
those in the chiral limit.
One of the low-energy constants at the one-loop order, $L_6$, appears for
this quantity.
The functions $\Delta(0,M^2)$ and $G(0,M^2,M^2)$ are given as
\begin{eqnarray}
  \label{eq:Delta}
  \Delta(0,M^2) &=& \frac{M^2}{16\pi^2} \ln\frac{M^2}{\mu_{sub}^2} +
                    g_1(M^2),
  \\
  \label{eq:G}
  G(0,M^2,M^2) &=& \frac{1}{2}\left[ \Delta(0,M^2)
                   + (M^2-M_\pi^2)\partial_{M^2}\Delta(0,M^2) \right].
\end{eqnarray}
They are evaluated at a ``pion mass'' as determined by the
Gell-Mann-Oakes-Renner (GMOR) relation
$M_{ij}^2=(m_i+m_j)\Sigma/F^2$, 
where the indices $i$ and $j$ label the sea quark mass or a fictitious
valence quark $v$.
For the sea quark mass, it gives a leading-order estimate of the
corresponding pion mass.
It slightly deviates from the actual pion mass calculated on the
lattice with the same quark mass, but the difference is from higher
orders of the chiral expansion and thus can be neglected at the order
considered for $\rho(\lambda)$.
The ``valence quark'' mass $m_v$ is taken at an imaginary value
$i\lambda$ to obtain the spectral function $\rho(\lambda)$ at a finite
$\lambda$, according to the procedure in \cite{Damgaard:2008zs}.
The scale parameter $\mu_{sub}$ denotes the renormalization scale, which
is conventionally taken at the $\rho$ meson mass.

The function $g_1(M^2)$ in (\ref{eq:Delta}) represents the finite
volume effect and is written in terms of a sum of the modified Bessel
function.
In the analysis of chiral extrapolation, we ignore the contribution of
$g_1(M^2)$, which is a good approximation for our data.
The largest possible finite volume effect may arise for the ensemble
of lightest pion with the smaller volume, {\it i.e.}
the $32^3\times 64$ lattice of $am_{ud}=0.0035$ at $\beta=4.17$, 
for which our estimate of $g_1(M^2)/F^2$ is $\sim$0.05 (0.02) at
$\lambda\simeq$ 5~MeV (10~MeV).
The maximum finite volume effect appears for smaller $\lambda$.
Even for this maximum case, the expected error due to neglecting such
effects is about the same size as the statistical error.
For the analysis of chiral extrapolation, we mainly use a larger bin
of size 15~MeV, for which the estimated finite volume effect is well
below the statistical error.

\begin{figure}[tbp]
  \centering
  \includegraphics[width=10cm,clip]{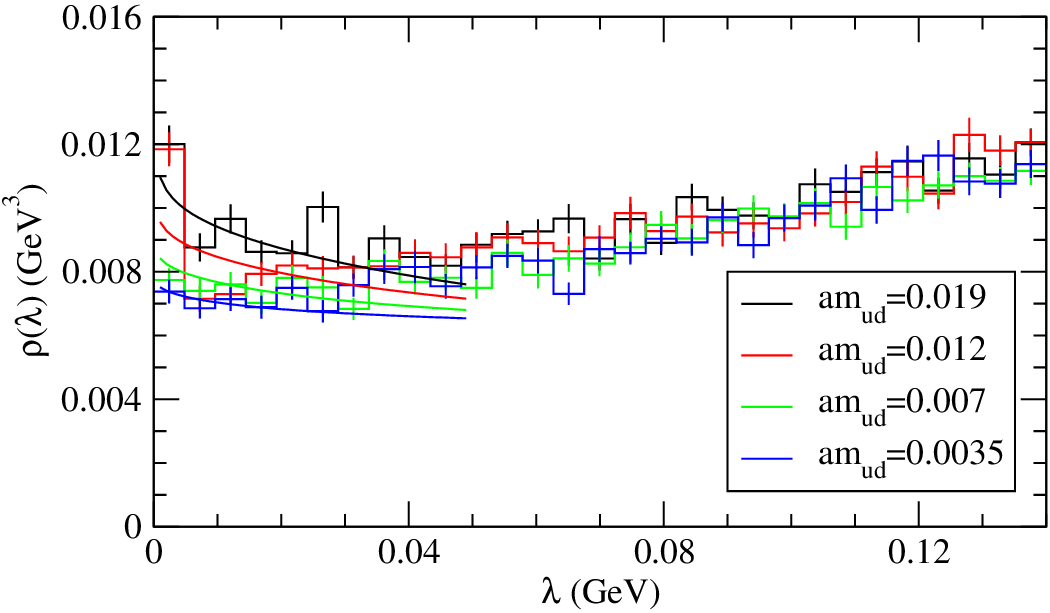}
  \includegraphics[width=10cm,clip]{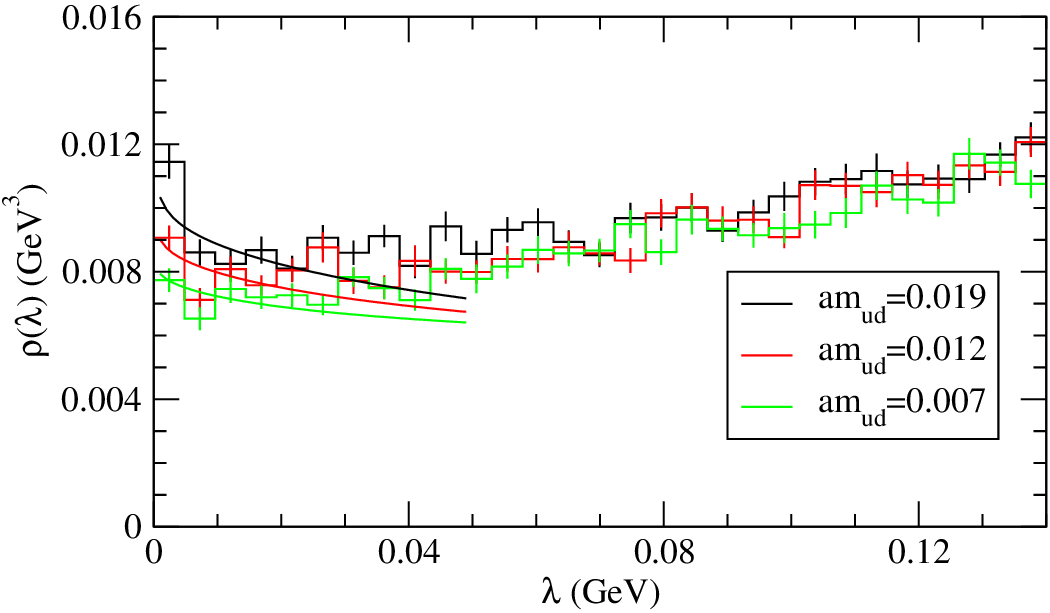}
  \caption{
    Dirac spectrum in the low-lying region.
    The data on the coarsest lattice ($\beta$ = 4.17) at two different
    strange quark masses: $am_s$ = 0.040 (top) and 0.030 (bottom).
    Results at $am_{ud}$ = 0.019 (black), 0.012 (red), 0.007 (green),
    0.0035 (blue) are plotted.
    Curves are from chiral perturbation theory. See text for details.
  }
  \label{fig:spect_b417}
\end{figure}

\begin{figure}[tbp]
  \centering
  \includegraphics[width=10cm,clip]{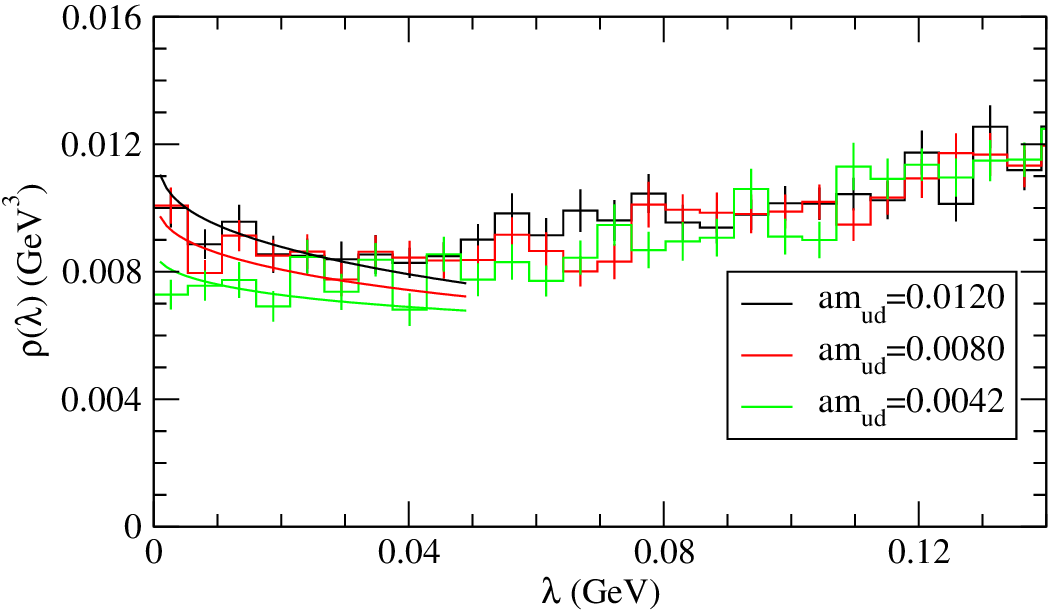}
  \includegraphics[width=10cm,clip]{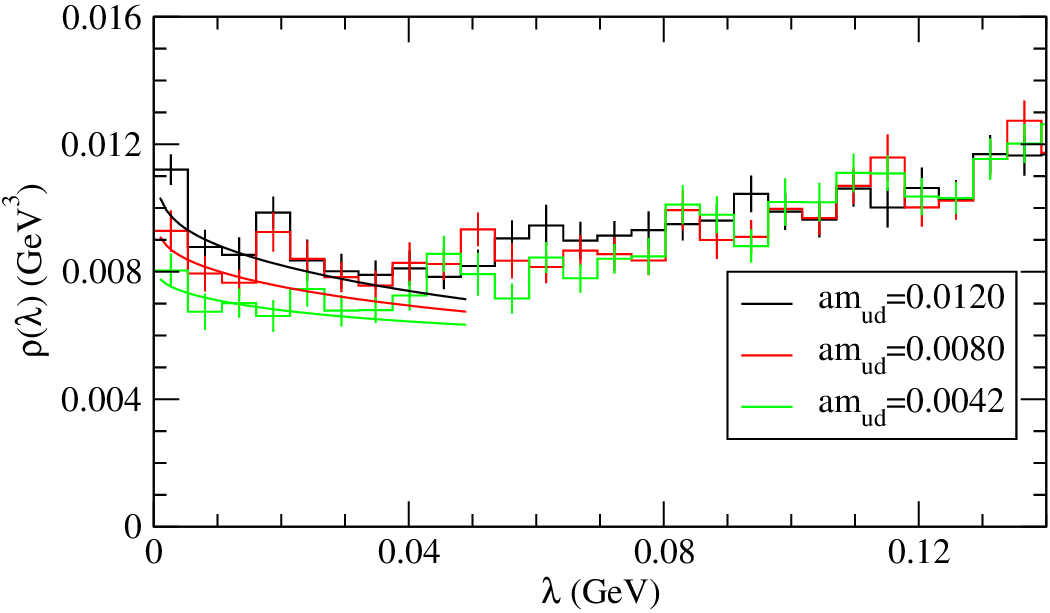}
  \caption{
    Dirac spectrum in the low-lying region.
    The data on the mid-fine lattice ($\beta$ = 4.35) at two different
    strange quark masses: $am_s$ = 0.025 (top) and 0.018 (bottom).
    Results at $am_{ud}$ = 0.0120 (black), 0.0080 (red), and 0.0045
    (green) are plotted.
    Curves are from chiral perturbation theory. See text for details.
  }
  \label{fig:spect_b435}
\end{figure}

\begin{figure}[tbp]
  \centering
  \includegraphics[width=10cm,clip]{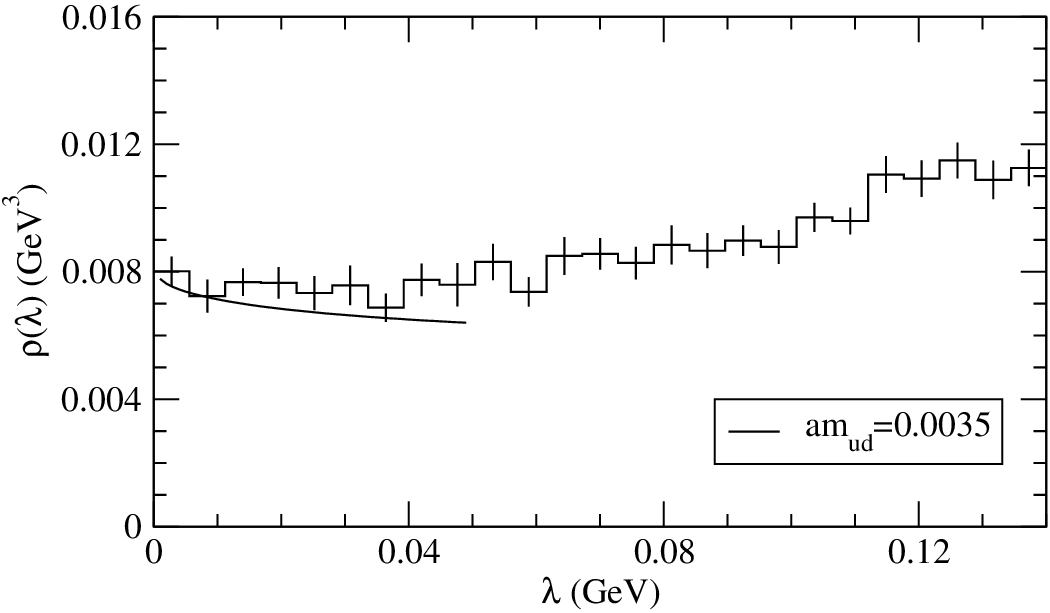}
  \caption{
    Dirac spectrum in the low-lying region.
    The data on the finest lattice ($\beta$ = 4.47) at 
    $am_{ud}$ = 0.0035 and $am_s$ = 0.0150 (black)
    are plotted.
    Curves are from chiral perturbation theory. See text for details.
  }
  \label{fig:spect_b447}
\end{figure}

In Figures~\ref{fig:spect_b417}--\ref{fig:spect_b447} 
we compare the lattice results with those of $N_f=2$ $\chi$PT 
at one-loop. 
The plots for each $\beta$ and strange quark mass are shown in
separate panels.
The lattice data are renormalized with the renormalization factor for
the scalar-density operator calculated separately using the
short-distance current correlator \cite{Tomii:2016xiv}.
The renormalization scheme is that of the $\overline{\mbox{MS}}$
scheme at the scale of 2~GeV.
The values are 
$Z_S(\mathrm{2~GeV})$ =
1.037(15),
0.934(9), and
0.893(7),
for $\beta$ = 4.17, 4.35, and 4.47, respectively.

From the data we can see a clear dependence on the up-down quark mass
near $\lambda=0$.
In $\chi$PT, the quark mass dependence is induced 
at the one-loop order through the functions $\Delta(0,M^2)$ and
$G(0,M^2,M^2)$ as well as through the counter term including $L_6$.
Another prominent feature of the low-mode spectrum $\rho(\lambda)$ is
the increase below $\lambda\sim$ 20~MeV, which is more pronounced for
heavier sea quarks, 
while the rise almost disappears at the lightest up and down sea
quarks available at $\beta$ = 4.17 
(upper panel of Figure~\ref{fig:spect_b417}).

The one-loop $\chi$PT prediction of $N_f=2$ is shown by curves in 
Figures~\ref{fig:spect_b417}--\ref{fig:spect_b447}.
The curves are for 
$\Sigma^{1/3}(\mathrm{2~GeV})$ = 270~MeV and $L_6$ = 0.0030,
which are the central values of a fit (see below) with a nominal value
of $F$ = 90~MeV.
The strange quark mass dependence is introduced assuming a linear
dependence of $\Sigma^{1/3}(\mathrm{2~GeV})$ on $m_s$.
The value of $\Sigma^{1/3}(\mathrm{2~GeV})$ mentioned is at the
physical strange quark mass.

Figures~\ref{fig:spect_b417}--\ref{fig:spect_b447} demonstrate that
the $\chi$PT curves also show the increase toward $\lambda=0$
especially for heavier sea quarks and nicely reproduce the lattice
data, which show the increase below $\lambda\sim$ 15--20~MeV.
This is not due to a tuning of parameters.
In fact, the extra parameter $L_6$ appearing at the one-loop order 
controls only the overall shift of $\rho(\lambda)$ without influencing
its $\lambda$ dependence.
The functional form of the pion-loop contribution, $\mathrm{Re}\Delta(0,M_{vi}^2)$ and 
$\mathrm{Re}G(0,M_{vv}^2,M_{vv}^2)$, is responsible for the increase
toward $\lambda$ = 0.
On the other hand, the one-loop $\chi$PT formula does not explain the
slight growth toward larger $\lambda$ above $\lambda\sim$ 20~MeV.
The higher order calculations would be needed to describe this regime.

With the $N_f=3$ $\chi$PT in which kaons and $\eta$ are also taken as
the dynamical degrees of freedom of chiral effective theory,
the number of parameters is reduced as we do not need to separately
model the strange quark mass dependence.
It turned out that a formula including $\Sigma^{1/3}(\mathrm{2~GeV})$,
$L_6$ and a parameter to describe the discretization effect as
fit parameters does not fit the data well.
($\chi^2$/dof is larger than 3.5.)
It is probably due to too large strange quark mass to be treated
within the $\chi$PT framework.
In fact, our data for $\rho(\lambda)$ deviates from the one-loop
$\chi$PT results above $\lambda\simeq$ 20~MeV.
The physical strange quark mass 90--100~MeV is far beyond this
threshold. 

We determine the parameters $\Sigma$ and $L_6$ through a fit of the
lattice data while fixing $F$ = 90~MeV.
The fit is done for the value of 
\begin{equation}
  \bar{\rho}[0:\delta] = \frac{1}{\delta}
  \int_0^\delta d\lambda\rho(\lambda)
\end{equation}
with $\delta$ = 0.015~GeV.
Both the lattice data and the $\chi$PT formula are integrated in the
region $[0,\delta]$.
This value of $\delta$ corresponds to $2\delta\Sigma/F^2\simeq$
250~MeV, which is well below the kaon mass.
It corresponds to the lowest three bins in the plots shown in 
Figures~\ref{fig:spect_b417}--\ref{fig:spect_b447}.
In this region, the $\chi$PT formula describes the data quite well.

The strange quark mass dependence of $\rho(\lambda)$ is introduced
assuming a linear dependence of $\rho(\lambda)$ on $m_s$.
In the narrow range of the strange quark quark mass adopted in our
simulation and with the mild dependence of $\rho(\lambda)$ on $m_s$,
this approximation should describe the data well.
Namely, we multiply 
\begin{equation}
  \label{eq:ms-dep}
  1+c_s(M_{\eta_{ss}}^2-M_{\eta_{ss}}^{\mathrm{(phys)}2})
\end{equation}
as an overall factor to $\rho(\lambda)$ in (\ref{eq:rho})
to interpolate the data to the physical strange quark mass.
The parameter $c_s$ is to be determined by a fit.
Here, $M_{\eta_{ss}}^{\mathrm{(phys)}}$ = 687~MeV is a mass of
fictitious $s\bar{s}$ pseudo-scalar meson estimated using the GMOR
relation. 
Our lattice ensembles contain those of different strange quark masses
while other parameters are fixed.
The strange quark masses in the simulations are chosen in such a way
that they sandwich its physical value.
We in effect interpolate between them by (\ref{eq:ms-dep}).

Similarly, the discretization effect is parameterized by a linear
function in $a^2$, multiplying $1+c_aa^2$ as an overall factor with
$c_a$ a fit parameter.

The fit for the all available data points yield
$\Sigma^{1/3}(\mathrm{2~GeV})$ =  270.0(1.3)~MeV,
$L_6$ = 0.00016(6), 
as well as
$c_s$ = 0.50(30)~GeV$^{-2}$,
$c_a$ = 0.00(15)~GeV$^2$,
with $\chi^2/\mathrm{dof}$ = 1.29.
As advertised, the discretization effect is invisible within the
statistical error.

\begin{figure}[tbp]
  \centering
  \includegraphics[width=10cm,clip]{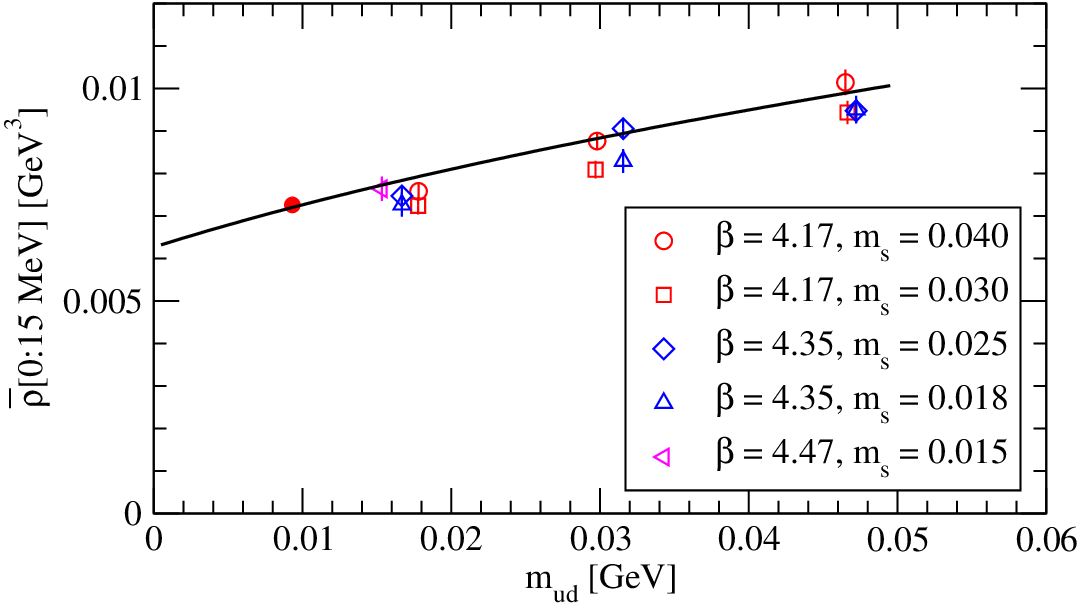}
  \caption{
    Chiral extrapolation of $\bar{\rho}[0:\delta]$.
    Data are averaged in the lowest bin of $\delta$ = 15~MeV.
    Open symbols are lattice data from each ensemble. 
    Different symbols correspond to different lattice spacings and
    strange quark masses, as denoted in the inset.
    A filled circle near $m_{ud}$ = 0.01~GeV is that of lightest quark
    at $\beta$ = 4.17 and $am_s$ = 0.040
    on the larger volume $48^3\times 96$.
    Data are plotted against the up-down quark mass $m_{ud}$
    renormalized with the $\overline{\mathrm{MS}}$ scheme at a scale
    2~GeV.
    The solid curve is that of the one-loop $\chi$PT fit in the
    continuum limit and at the physical strange quark mass.
  }
  \label{fig:chext}
\end{figure}

Chiral extrapolation of $\bar{\rho}[0:\delta]$ is shown in 
Figure~\ref{fig:chext} as a function of sea up and down quark mass 
$m_{ud}$.
Data points do not lie on a single universal curve because the data at
different strange quark masses are put in the same plot.
In other words, there is a significant strange quark mass dependence,
which seems to be well described by an overall shift of the curve.
Dependence on the lattice spacing is not very significant from the
plot, as the fit also suggests.
The curvature due to the one-loop correction is not strong but still
visible, and makes the chiral limit slightly lower than a naive linear
extrapolation in $m_{ud}$.

We list the possible sources of systematic errors in the following.
First of all, the renormalization constant $Z_S(\mathrm{2~GeV})$
determined in \cite{Tomii:2016xiv} contains some errors.
(The numbers are given above.)
The size is 1.4\%, 1.0\% and 0.8\% for coarse, medium and fine
lattices, respectively.
We take the largest error, 1.4\%, to be conservative, for the estimate
of the error for $\Sigma(\mathrm{2~GeV})$.
When we quote the number for $\Sigma^{1/3}(\mathrm{2~GeV})$, we
therefore assign 0.5\% as an estimated systematic error from this
source. 

The discretization effect is well under control in our calculation.
In fact, our fit implies that the lattice-spacing dependence is
consistent with zero.
Although it is insignificant, by keeping the term describing this
effect in the fit function, we can take account of possible systematic
effect.
We therefore do not add extra errors from the discretization effects.

Finite volume effect is explicitly checked on the ensembles with the 
lightest pion ($\sim$ 230~MeV) as shown in
Figure~\ref{fig:spect_b417_vol}.
We do not observe any statistically significant difference between the 
two volumes ($32^3$ and $48^3$), which is consistent with an
expectation from $\chi$PT, {\it i.e.} the predicted size of the finite
volume effect is about 5\% for the smaller lattice and is about the
same size as the statistical error.
For heavier pions the $\chi$PT predicts exponentially suppressed
finite volume effects.
Therefore, for all the data used in the fit to extract the chiral
condensate, this source of error is within our statistical error.
(Note that the smaller volume data at $M_\pi\sim$ 230~MeV are not
included in the fit.)

Higher-order corrections from $\chi$PT may be significant especially
for larger $\lambda$ and heavier quarks.
Since we can explicitly confirm the consistency of the lattice data
with the one-loop $\chi$PT for its $\lambda$-dependence in the range
of our analysis, we expect that two-loop correction is insignificant
below $\lambda$ = 15~MeV.
We checked that the result with a slightly smaller bin size,
10~MeV, is consistent within the statistical error.
Also for the quark mass, the one-loop $\chi$PT fits the lattice data
well up to the data points of heaviest pion masses ($\sim$ 500~MeV).
In order to examine the significance of the higher order effects, 
we tried to fit the data with a function including the analytic
terms of $O(M^4/F^4)$. 
The coefficient obtained from such an analysis is of order of 
$3\times 10^{-6}$ and statistically consistent with zero.
The best fit value of $\Sigma^{1/3}(\mathrm{2~GeV})$ is shifted by
only 0.1~MeV, which is much smaller than the statistical error.
We can conclude that such effects are well below the statistical error
in our analysis.

There is a potential effect of slightly inaccurate
implementation of the Ginsparg-Wilson relation with the M\"obius
domain-wall fermion.
As we already discussed, the 4D effective operator of the M\"obius
domain-wall fermion violates the Ginsparg-Wilson relation by the
amount characterized by the residual mass, which is about 1~MeV on our
coarsest lattice and an order of magnitude smaller on finer lattices.
It means that the eigenvalue of the Dirac operator is distorted by the
amount of $O(\mathrm{1~MeV})$ on the lattices at $\beta$ = 4.17.
Since the bin size in the analysis is much larger (= 15~MeV), the
error due to this effect is minor.
Moreover, the effect should be negligible on finer lattices, and it is
also taken into account by the continuum extrapolation.
We therefore do not introduce additional error budget for this
effect. 

Finally, our input value for lattice spacing has an error of 1.7\%, which
affects dimensionful quantities, including the chiral condensate.
We therefore add this size of error for $\Sigma^{1/3}$.

Having these various systematic errors considered, we quote
\begin{equation}
  \label{eq:result}
  \Sigma^{1/3}(\mathrm{2~GeV}) = \mbox{270.0(1.3)(1.3)(4.6)~MeV},
\end{equation}
where the errors are those from statistical, renormalization, and
lattice scale, respectively.
Adding in quadrature, the total error is 4.9~MeV, which is 1.8\%.
The Flavour Lattice Averaging Group (FLAG) quotes the chiral
condensate for $N_f$ = 2+1,
$\Sigma^{1/3}(\mathrm{2~GeV})$ = 274(3)~MeV \cite{Aoki:2016frl},
as an average of
\cite{Blum:2014tka,Bazavov:2010yq,Borsanyi:2012zv,Durr:2013goa}.
They are obtained by fitting meson masses and decay constants with the
$\chi$PT formulae, where the chiral condensate appears as a
coefficient in the Gell-Mann-Oakes-Renner (GMOR) relation.
Our result (\ref{eq:result}) is consistent with the world average and
the precision is comparable.

\section{Conclusion}
\label{sec:conclusion}
The eigenvalue spectrum of the Dirac operator reflects the quantum
effects of QCD.
The near-zero eigenvalue regime is special, as it can be connected to
the order parameter of spontaneous chiral symmetry breaking in QCD,
{\it i.e.} the chiral condensate.
This relation known as the Banks-Casher relation can be extended to
the case of finite $\lambda$ as well as finite quark masses using
$\chi$PT.
This work provides a direct test of these relations by calculating the
spectral function in lattice QCD simulations.

The M\"obius domain-wall fermion formulation used in this work to
define the Dirac operator possesses an approximate chiral symmetry
with an error of order 1~MeV at most, and the accumulation of the
eigenvalues above this value is not much affected by this artifact.
We extract the chiral condensate from the spectrum below 15~MeV by
fitting the lattice data with the $\chi$PT formula.
The discretization error is well under control and even extrapolated
away to the continuum limit using relatively fine-grained lattices of
$a$ = 0.080--0.044~fm.

The remaining uncertainty is at the level of 2\% for $\Sigma^{1/3}(\mathrm{2~GeV)}$.
This provides a precise test of the GMOR relation,
since there is no free parameter left for the leading-order equation
$m_\pi^2/m=2\Sigma/F^2$ once $m_\pi$ and $F$ are calculated.
The agreement of our result 270.0(4.9)~MeV with that of an average of
previous results obtained through GMOR gives further evidence
supporting $\chi$PT as an effective theory of QCD at low energies.

The eigenvalue filtering technique utilized in this work is proven to
be effective to obtain the spectral function of the Dirac operator.
In this analysis we used only the near-zero regime of the eigenvalues,
while the entire spectrum is calculated as a by-product.
Such information may be useful to extract the mass anomalous dimension
of QCD with a non-perturbative method as discussed in
\cite{Fodor:2016hke}.

\section*{Acknowledgment}
We are grateful to Julius Kuti for fruitful discussions and in
particular for bringing our attention to the method introduced in 
\cite{Fodor:2016hke}.
We thank other members of the JLQCD collaboration.
This work is a part of its research programs.
Numerical simulations are performed on Hitachi SR16000 and IBM Blue
Gene/Q systems at KEK under its Large Scale Simulation Program
(No. 15/16-09).
This work is supported in part 
by JSPS KAKENHI Grant Numbers JP25800147, JP26247043 and JP26400259, and
by the Post-K supercomputer project through JICFuS.



\begin{thebibliography}{99}

\bibitem{Banks:1979yr}
  T.~Banks and A.~Casher,
  Nucl.\ Phys.\ B {\bf 169} (1980) 103.
  doi:10.1016/0550-3213(80)90255-2

\bibitem{Smilga:1993in} 
  A.~V.~Smilga and J.~Stern,
  Phys.\ Lett.\ B {\bf 318}, 531 (1993).
  doi:10.1016/0370-2693(93)91551-W

\bibitem{Osborn:1998qb}
  J.~C.~Osborn, D.~Toublan and J.~J.~M.~Verbaarschot,
  Nucl.\ Phys.\ B {\bf 540}, 317 (1999)
  doi:10.1016/S0550-3213(98)00716-0
  [hep-th/9806110].

\bibitem{Akemann:2016keq}
  G.~Akemann,
  arXiv:1603.06011 [math-ph].

\bibitem{Damgaard:2008zs} 
  P.~H.~Damgaard and H.~Fukaya,
  JHEP {\bf 0901}, 052 (2009)
  [arXiv:0812.2797 [hep-lat]].

\bibitem{Giusti:2008vb} 
  L.~Giusti and M.~Luscher,
  JHEP {\bf 0903}, 013 (2009)
  [arXiv:0812.3638 [hep-lat]].

\bibitem{Fukaya:2007fb} 
  H.~Fukaya {\it et al.} [JLQCD Collaboration],
  Phys.\ Rev.\ Lett.\  {\bf 98}, 172001 (2007)
  doi:10.1103/PhysRevLett.98.172001
  [hep-lat/0702003].

\bibitem{Fukaya:2007yv} 
  H.~Fukaya {\it et al.} [TWQCD Collaboration],
  Phys.\ Rev.\ D {\bf 76}, 054503 (2007)
  doi:10.1103/PhysRevD.76.054503
  [arXiv:0705.3322 [hep-lat]].

\bibitem{Fukaya:2009fh} 
  H.~Fukaya {\it et al.} [JLQCD Collaboration],
  Phys.\ Rev.\ Lett.\  {\bf 104}, 122002 (2010)
  Erratum: [Phys.\ Rev.\ Lett.\  {\bf 105}, 159901 (2010)]
  doi:10.1103/PhysRevLett.104.122002, 10.1103/PhysRevLett.105.159901
  [arXiv:0911.5555 [hep-lat]].

\bibitem{Fukaya:2010na} 
  H.~Fukaya {\it et al.} [JLQCD and TWQCD Collaborations],
  Phys.\ Rev.\ D {\bf 83}, 074501 (2011)
  doi:10.1103/PhysRevD.83.074501
  [arXiv:1012.4052 [hep-lat]].

\bibitem{Damgaard:1997ye}
  P.~H.~Damgaard and S.~M.~Nishigaki,
  Nucl.\ Phys.\ B {\bf 518}, 495 (1998)
  doi:10.1016/S0550-3213(98)00123-0
  [hep-th/9711023].

\bibitem{Damgaard:2000ah}
  P.~H.~Damgaard and S.~M.~Nishigaki,
  Phys.\ Rev.\ D {\bf 63}, 045012 (2001)
  doi:10.1103/PhysRevD.63.045012
  [hep-th/0006111].

\bibitem{Engel:2014cka} 
  G.~P.~Engel, L.~Giusti, S.~Lottini and R.~Sommer,
  Phys.\ Rev.\ Lett.\  {\bf 114}, no. 11, 112001 (2015)
  doi:10.1103/PhysRevLett.114.112001
  [arXiv:1406.4987 [hep-ph]].

\bibitem{Engel:2014eea} 
  G.~P.~Engel, L.~Giusti, S.~Lottini and R.~Sommer,
  Phys.\ Rev.\ D {\bf 91}, no. 5, 054505 (2015)
  doi:10.1103/PhysRevD.91.054505
  [arXiv:1411.6386 [hep-lat]].

\bibitem{Cichy:2013gja} 
  K.~Cichy, E.~Garcia-Ramos and K.~Jansen,
  JHEP {\bf 1310}, 175 (2013)
  doi:10.1007/JHEP10(2013)175
  [arXiv:1303.1954 [hep-lat]].

\bibitem{DiNapole:2013}
  E.~Di~Napoli, E.~Polizzi and Y.~Saad,
  arXiv:1308.4275 [cs.NA].

\bibitem{Ginsparg:1981bj} 
  P.~H.~Ginsparg and K.~G.~Wilson,
  Phys.\ Rev.\ D {\bf 25}, 2649 (1982).
  doi:10.1103/PhysRevD.25.2649

\bibitem{Cossu:2016yzp} 
  G.~Cossu, H.~Fukaya, S.~Hashimoto, T.~Kaneko and J.~Noaki,
  arXiv:1601.00744 [hep-lat].

\bibitem{Fodor:2016hke} 
  Z.~Fodor, K.~Holland, J.~Kuti, S.~Mondal, D.~Nogradi and C.~H.~Wong,
  arXiv:1605.08091 [hep-lat].

\bibitem{Brower:2012vk} 
  R.~C.~Brower, H.~Neff and K.~Orginos,
  arXiv:1206.5214 [hep-lat].

\bibitem{Kaplan:1992bt} 
  D.~B.~Kaplan,
  Phys.\ Lett.\ B {\bf 288}, 342 (1992)
  doi:10.1016/0370-2693(92)91112-M
  [hep-lat/9206013].

\bibitem{Shamir:1993zy} 
  Y.~Shamir,
  Nucl.\ Phys.\ B {\bf 406}, 90 (1993)
  doi:10.1016/0550-3213(93)90162-I
  [hep-lat/9303005].

\bibitem{Noaki:2014sda} 
  J.~Noaki {\it et al.} [JLQCD Collaboration],
  PoS LATTICE {\bf 2014}, 069 (2014).

\bibitem{Morningstar:2003gk} 
  C.~Morningstar and M.~J.~Peardon,
  Phys.\ Rev.\ D {\bf 69}, 054501 (2004)
  doi:10.1103/PhysRevD.69.054501
  [hep-lat/0311018].

\bibitem{Borsanyi:2012zs} 
  S.~Borsanyi {\it et al.},
  JHEP {\bf 1209}, 010 (2012)
  doi:10.1007/JHEP09(2012)010
  [arXiv:1203.4469 [hep-lat]].

\bibitem{Hashimoto:2014gta} 
  S.~Hashimoto, S.~Aoki, G.~Cossu, H.~Fukaya, T.~Kaneko, J.~Noaki and P.~A.~Boyle,
  PoS LATTICE {\bf 2013}, 431 (2014).

\bibitem{Tomii:2016xiv} 
  M.~Tomii {\it et al.} [JLQCD Collaboration],
  arXiv:1604.08702 [hep-lat].

\bibitem{Nakayama:2016atf} 
  K.~Nakayama, B.~Fahy and S.~Hashimoto,
  arXiv:1606.01002 [hep-lat].

\bibitem{Fukaya:2015ara} 
  H.~Fukaya {\it et al.} [JLQCD Collaboration],
  Phys.\ Rev.\ D {\bf 92}, no. 11, 111501 (2015)
  doi:10.1103/PhysRevD.92.111501
  [arXiv:1509.00944 [hep-lat]].

\bibitem{Fahy:2015xka} 
  B.~Fahy, G.~Cossu, S.~Hashimoto, T.~Kaneko, J.~Noaki and M.~Tomii,
  arXiv:1512.08599 [hep-lat].

\bibitem{Cossu:2013ola} 
  G.~Cossu, J.~Noaki, S.~Hashimoto, T.~Kaneko, H.~Fukaya, P.~A.~Boyle and J.~Doi,
  arXiv:1311.0084 [hep-lat].

\bibitem{Aoki:2016frl} 
  S.~Aoki {\it et al.},
  arXiv:1607.00299 [hep-lat].

\bibitem{Blum:2014tka} 
  T.~Blum {\it et al.} [RBC and UKQCD Collaborations],
  Phys.\ Rev.\ D {\bf 93}, no. 7, 074505 (2016)
  doi:10.1103/PhysRevD.93.074505
  [arXiv:1411.7017 [hep-lat]].

\bibitem{Bazavov:2010yq} 
  A.~Bazavov {\it et al.},
  PoS LATTICE {\bf 2010}, 083 (2010)
  [arXiv:1011.1792 [hep-lat]].

\bibitem{Borsanyi:2012zv} 
  S.~Borsanyi, S.~Durr, Z.~Fodor, S.~Krieg, A.~Schafer, E.~E.~Scholz and K.~K.~Szabo,
  Phys.\ Rev.\ D {\bf 88}, 014513 (2013)
  doi:10.1103/PhysRevD.88.014513
  [arXiv:1205.0788 [hep-lat]].

\bibitem{Durr:2013goa} 
  S.~D\"urr {\it et al.} [Budapest-Marseille-Wuppertal Collaboration],
  Phys.\ Rev.\ D {\bf 90}, no. 11, 114504 (2014)
  doi:10.1103/PhysRevD.90.114504
  [arXiv:1310.3626 [hep-lat]].

\end{thebibliography}
\end{document}